\documentstyle[11pt,aaspp4]{article}
\def\mpc{\,h^{-1}{\rm Mpc}}
\def\kms{\,{\rm {km\, s^{-1}}}}
\begin{document}

\title{A Possible Bias Model for Quasars}

\author{Li-Zhi Fang\footnote{fanglz@time.physics.arizona.edu}
}

\affil{Department of Physics, University of Arizona, Tucson,
AZ 85721}

\and
\author{Y.P. Jing
  \footnote{jing@utaphp1.phys.s.u-tokyo.ac.jp}}

\affil{Department of Physics and Research Center for Early Universe
 (RESCEU), \\
School of Science, The University of Tokyo, Tokyo 113, Japan}

\begin{abstract}

  We propose that the majority of quasars at redshift $z\sim 1 - 5$
  formed in the environment of new born collapsed halos with 1-D
  velocity dispersion $\sigma_v^{1d} \sim 400 \kms$. The harboring
  coefficient $f$ of quasars per halo and the lifetime of quasars
  depend only on local process, not modulated by the density
  inhomogeneities on scales larger than the size of the halos. Thus,
  the bias of quasars on scale larger than the size of these halos is
  mainly determined by the parameter $\sigma_v$ used for quasar
  environment identification. With this model, the popular structure
  formation models, like SCDM and LCDM, can be fairly well reconciled
  with the data of quasars, including a. observed feature of the
  environment for quasars; b. redshift evolution of quasar abundance;
  c. the two-point correlation functions of quasars. This bias model
  predicts that the correlation function of quasars doesn't
  significantly evolve, or only slightly increases with redshift.
 
\end{abstract}


\section{Introduction}

\bigskip

Mass distribution at high redshifts is being a hot subject of the
large scale structure study. Data of various objects at moderate and
high redshifts, in particular, clusters of galaxies, are becoming
available for probing the formation and evolution of structures and
for discriminating among popular dark matter models (e.g. Jing \& Fang
1994; Eke, Cole \& Frenk 1996; Bahcall, Fan \& Cen 1997; Kitayama \&
Suto 1997). Considering that quasars are the most distant among various
luminous objects, they have also been applied in this approach
(e.g. Bi \& Fang 1997).

However, as a mass tracer of the cosmic matter field, quasars are
still playing a role different from clusters of galaxies. The problem
stems from so-called ``bias''. Clusters are a biased tracer of the
mass distribution. The correlation amplitude of clusters is believed
to be much higher than that of the underlying matter and increases
strongly with the cluster richness (Bahcall \& Soneira 1983). This
bias is plausibly explained by the mechanism that the observed
clusters are identified as massive collapsed halos of the density
field (Kaiser 1984).  That is, the bias of clusters is completely
determined by the gravitational parameters, like mass $M$ and virial
radius $r_{vir}$ used to identify the halos.  With this approach, a
detailed confrontation can be made between theories and the
observations of clusters.

Quasars may also be biased tracer of the mass distribution. Recent
observations indicate that the correlation amplitude of quasars may also
be different from the underlying dark matter (Mo \& Fang
1993; Komberg, Kravtsov \& Lukash 1994; Croom \& Shanks 1996; Franca,
Andreani \& Cristiani, 1997).  However, so far no detailed model is
available for the bias of quasars, though their high clustering
strength and environment imply that quasars are hosted by massive halos
(see below). Because of the lack of such a model, one cannot confront
the data of quasars with theoretical models in the way as for
clusters. For instance, the abundance of quasars can only be used as an
upper or lower limit to certain massive halos; no detailed comparison
between the number densities of quasars and of halos is
allowed. Obviously, it is very important to understand what kind of
mass halos are associated with the majority of quasars. Such a
knowledge will not only enable the observational data of quasars to be
powerful tests for theoretical models of galaxy formation but also
tell that what type of local environments is responsible for
intriguing the nuclear activities of quasars.

Like clusters and groups of galaxies, it is generally believed that
quasars should be associated with certain type of collapsed dark
matter halos. Yet, different from identification of clusters, the
environment suitable for forming quasars is not merely determined by
gravitational parameters, as the hydro processes are also
involved. Therefore, the identification of quasar-harboring halos
should be given by both gravitational and hydro parameters. In other
words, not all halos with certain $M$ and/or velocity dispersion
$\sigma_v$ harbor quasars, because certain hydro conditions must be
satisfied. However, considering the hydro processes are local, it is
reasonable to assume that the hydro conditions may not be modulated by
the density inhomogeneities on scales much larger than the size of the
halo $l$. In this case, the probability for a halo to have a quasar
should be the same for all halos of the same kind, without depending
on structures larger than $l$. Thus, the relative fractions of quasars
with respect to the certain collapsed halos should be the same for 
all volumes larger than $l^3$. Consequently, when
averaged on scale larger than $l$, the distribution of quasars
$n_{qso}({\bf r},z)$ at redshift $z$ should be proportional to that of
the considered halos, $n_{halo}({\bf r},z)$. Thus, all effects of the 
hydro processes can be absorbed into a ``normalization factor" $A$,
i.e. $n_{qso}({\bf r},z)=An_{halo}({\bf r},z)$, and $A$ is
less dependent on $z$ than $n_{halo}({\bf r},z)$. The bias of
quasar distribution with respect to the mass distribution is then
dominated by the bias of the selected halos with respect to the
mass. Based on this analysis, quasar bias, at least on large scales,
may also be only gravitational, depending on the gravitational
parameters used for selecting the quasar-suitable halos.

Accordingly, a possible model for quasar bias should at least satisfy
the three conditions. 1. Gravitational environment given by the
identified halos is consistent with the observed environment around
quasars; 2. The abundance of quasars, $n_{qso}(z)$, at redshift $z$ is
proportional to the number density of the identified halos,
$n_{halo}(z)$ in a redshift-independent way,
i.e. $n_{qso}(z)=An_{halo}$ where $A$ is a {\it z-independent}
constant, 3. The amplitude and $z$-evolution of the halo-halo
correlation function are consistent with the observed correlation
function of quasars. In this letter, we will show within the framework
of the CDM cosmogonic theories that such a bias model can indeed be
settled following the above-mentioned points.  The details of the
points 1, 2 and 3 will be discussed in the \S 2, 3 and 4,
respectively.

\section{Basic assumption: The gravitational condition for a quasar halo}

A bias model of quasars is just a phenomenological relationship
between the cosmic density field and quasars. With this relationship
quasars can be identified from the mass density field. In this sense,
bias, in fact, is a model for the environment suitable for the quasar
formation.

In what environment are quasars most likely to be formed? Many 
observations indicate that quasars are preferentially located in
groups of galaxies. The evidences include the quasar-galaxy covariance
function (Yee \& Green 1987), the galaxy environments around quasars
(Ellingson, et al. 1991), clustering of quasars (Bahcall \& Chokshi
1991; Mo \& Fang 1993; Komberg, Kravtsov \& Lukash 1994; Croom \&
Shanks 1996; Franca, Andreani \& Cristiani, 1997), and the
CIV-associated absorption in high redshift radio-loud quasars (Foltz
et al 1988).   Recently, an
observation of a companion to quasar BR1202-0725 with high redshift
$z=4.7$ directly shows that the width of their Ly$\alpha$ emission
lines is around 400 km s$^{-1}$ (Petitjean et al. 1996.). These
observations seemingly point to quasars being identified with the 
newly collapsed halos with 1-D velocity dispersion like
groups, i.e. $\sigma_v^{1d}\approx
400\kms$. It should be pointed out that this environment condition may
not be necessary for low redshift quasars (Smith, Boyle \& Maddox, 1995),
because galaxies and clusters underwent a significant evolution at 
$z \sim 0.5$.  But, it is reasonable to assume that the environment with
$\sigma_v^{1d}\approx 400\kms$ is favored by the formation of quasars
at higher redshifts: a) enough collapsed objects to form the engine of a
quasar; b) dispersed gas to feed the engine through accretion; and c)
not {\it too} many proto-galaxies to possibly disrupt the process of
quasar formation. In the next two
sections, we shall compare, within the framework of CDM cosmogonic
models, the theoretical predictions for such halos with the observed
quasar abundance and correlation functions.

\section{Redshift Evolution of Abundance of quasars}

 For a Gaussian initial perturbation, the comoving number density of
halos with 1-D velocity dispersion $\sigma_v^{1d}$ can be calculated
with the Press \& Schechter formalism (1974) as
\begin{eqnarray}
\lefteqn{n(\sigma_v^{1d},z) d\sigma_v^{1d}=
  -{\sqrt{3} \over (2\pi)^{3/2} R^3} } \\  \nonumber
& & \times {d\ln \Delta(R,z)\over d \ln R} {\delta_c \over \Delta(R,z)} 
 \exp \left(-{\delta_c^2 \over \Delta^2(R,z)} \right)
{dR\over R}\,
\end{eqnarray}
where $R$ is the Lagrangian radius of the dark halo being considered
and $\delta_c\approx1.69$ almost independent of cosmologies.
$\Delta(R,z)$ is the $rms$ of the linear density fluctuations at
redshift $z$ within a top-hat window of radius $R$, and is determined
by the initial density spectrum $P(k)$ and normalization factor
$\sigma_{8}=\Delta(8\mpc,0)$.  The relationship between
$\sigma_v^{1d}$ and $R$ is given by (Narayan \& White 1988)
\begin{equation}
\sigma_v^{1d}= c_\sigma \Omega_0^{1/2} H_0 R_0
(1+z)^{1/2} .
\end{equation}
for a universe with the density parameter $\Omega_0$. The N-body
simulation results (Jing \& Fang 1994) showed $c_\sigma \approx 1.2$
which is used for all calculations in this work.  Two representative
CDM models, i.e. the standard CDM (SCDM) and flat low-density CDM
(LCDM) models are employed. The Hubble constant $h = H_0/100 \kms$
Mpc$^{-1}$, mass density $\Omega_0$, cosmological constant $\lambda_0$
and $\sigma_{8}$ are taken to be $(0.5, 1, 0, 0.58)$ and $(0.75, 0.3,
0.7, 1)$ for the SCDM and LCDM, respectively. The models with these
parameters seem to provide a good approximation to many observational
properties of the Universe, especially the abundance of clusters
(which is much related to the topic of this work).

The total number density of the collapsed halos with the velocity
dispersion greater than a certain value, say $\sigma_{lim}^{1d}$, is
\begin{equation}
N(>\sigma^{1d}_{lim}, z)=\int^{\infty}_{\sigma^{1d}_{lim}}
n(\sigma^{1d}_v,z)d\sigma^{1d}_v.
\end{equation}
The birth rate of halos with $\sigma^{1d}_v \geq \sigma^{1d}_{lim}$ is
$dN(>\sigma^{1d}_{lim},z)/dt$. This birth rate is shown in Fig.1 for
$\sigma^{1d}_{lim}=$ 200, 400, and 800 $\kms$.  For each
$\sigma^{1d}_{lim}$, the birth rate in the two models possesses
similar shape. At $\sigma^{1d}_{lim}=400\kms$, the birth rate keeps
steady from $z=5$ to about 2, and then rapidly drops to zero at
$z\sim 0.3$ for SCDM and 0.7 for LCDM. The peak birth rate however
shifts to higher redshift for smaller $\sigma^{1d}_{lim}$, and to
lower redshift for larger $\sigma^{1d}_{lim}$.  As has been discussed
in \S 1 and 2, each newly collapsed halo with $\sigma^{1d}_{v} >
\sigma^{1d}_{lim}$ may host $f$ quasars in average, and the harboring
coefficient $f$ is $z$-independent. If the mean lifetime of quasars is
$t_{qso}$, which is also $z$-independent, the comoving number density
of quasars at the epoch of redshift $z$ is given by
\begin{equation}
n(z) = f\int_{t(z)-t_{qso}}^{t(z)}
\frac{\partial N(>\sigma^{1d}_{lim}, z')}{\partial z'}\frac{d z}{dt} dt.
\end{equation}
An accurate quasar lifetime $t_{qso}$ is not important 
only if it is $z$-independent and is much less
than the age of the universe
for redshift $z \leq 5$. Considering that birth rate of the halos is
slowly varying with redshift when
$z > 1$, we have approximately $n(z) \simeq ft_{qso} [\partial
N(>\sigma^{1d}_{lim}, z)/\partial z] (dz/dt)$. Since the shape of the
birth rate is a strong function of the velocity dispersion (especially 
as a function of $z$), the abundance of quasars provides a strong test 
to the bias model proposed here.

 Fig.2 plots $n(z)$ for the SCDM and LCDM models. The ``normalization
constant" $t_{qso} f$ are adjusted in order that the theoretical
maximum number densities of quasar abundance can fit with the observed
one.  The data points are the number density of quasars complete to
absolute magnitude $M_B = -26$ (Pei, 1995, Hewett, Foltz \& Chaffee
1993, Schmidt, Schneider \& Gunn 1992). In these observations, the
density $n(z)$ is measured in the Einstein - de Sitter cosmology
($\Omega=1$, $\lambda_0=0$ and $h=0.5$). These data have been
corrected to the case of non-zero $\lambda_0$ in the panel of
LCDM. The figure shows the redshift evolution of quasar abundances is
fairly well described by the newly collapsed halos with
$\sigma^{1d}_{lim}=400\kms$. Both the SCDM and LCDM models are in
reasonably good agreement with the observed abundance $n_{qso}(z)$ if
the {\it constant} ($z$-independent) parameter $t_{qso} f$ is taken to
be $0.02\cdot 10^7$ yr for the SCDM model and $0.33\cdot 10^7$ yr for
the LCDM model, though the best fitting value of $\sigma_v^{1d}$ is
slightly lower (about $340\kms$) for the SCDM model.  However for
$\sigma^{1d}_{lim} =200 $, and 800 $\kms$, the predictions cannot fit
with observed redshift evolution regardless of how to adjust the
``normalization constant" $t_{qso} f$. Therefore, the consistency
between theoretical and observed redshift evolution of quasar
abundance can be achieved only for $\sigma^{1d}_v\approx 400 \kms$
halos, which is not trivial. This result strongly supports that the
majority of quasars be associated with new born halos with 1-D
$\sigma^{1d}_{lim} \sim 400 \kms$.

Particularly, for an environment with a given mass or velocity
dispersion, the luminosities and lifetimes of quasars are still
dispersed. However, if the distributions of the luminosities and of
the lifetimes for this type of halos are not modulated by large scale
perturbations, all conclusion here should hold.

\section{Amplitude and redshift-evolution of quasar correlation function}

According to the analysis of \S 1, the bias model for
$\sigma^{1d}_{lim} \sim 400$ km s$^{-1}$ halos is available on the scales
larger than their typical size 
which is about $3(1+z)^{-1/2}$ $\Omega_0^{-1/2} h^{-1}$
Mpc. Therefore, on scales larger than 5 $h^{-1}$ Mpc the bias of
quasar distribution is due mainly to the bias of $\sigma^{1d}_v$
halos. On such large scales and at high redshifts, the mass distribution
is still linear. Therefore, the two-point correlation function of the
halos at redshift $z$ is given by
\begin{equation}
\xi(r,z) = b^2(R,z) \xi_m(r,z)
\end{equation}
where $r$ is physical radius, $\xi_m(r,z)$ is the linear mass
correlation function, and the bias factor $b$ is given by
(Mo \& White 1996)
\begin{equation}
b(R,z) = 1 + \frac{1}{\delta_c}
  \left [ \frac{\delta_c^2}{\Delta^2(R,z)} - 1 \right].
\end{equation}

 Fig.3a shows the correlation functions for $\sigma^{1d}_{lim}=200$, 400
and 800 $\kms$ halos at $z =2$. It shows that for
$\sigma^{1d}_{lim}=400 \kms$ the correlation length $r_0$ defined by
$\xi(r)=(r/r_0)^{-1.8}$ is $\sim 6$ h$^{-1}$ Mpc for the SCDM, and
$\sim 14$ h$^{-1}$ Mpc for the LCDM. Observationally, the two-point
correlation function of quasars is found to obey the same power
law. With $q_0$ taken to be 0.5, the correlation length $r_0$ at
$z=1.5$ is found to be $\sim 6.6 \pm 0.5 \ h^{-1}$ Mpc (Mo \& Fang
1993), $10 \pm 2 \ h^{-1}$ Mpc (Komberg, Kravtsov \& Lukash 1994), and
$6.2 - 8.0 \ h^{-1}$ Mpc (Franca, Andreani \& Cristiani 1997). Actually, 
these results do not sensitively depend on $q_0$, because on the length 
scales of about $10 \ h^{-1}$ Mpc, the influence of the parameter
$q_{0}$ is small. We should be careful in comparing the observed data with
 Fig.3a. Generally, the samples of quasars employed for the correlation
statistics possess different limit magnitude for different redshifts. The
samples employed for the correlation statistics contain quasars
with $M_b > -26$. Namely, they are not complete in the sense as the data
used in Fig.2. The larger $M_B$ may lead to a lower amplitude
of the correlations. Considering this uncertainty, both SCDM and LCDM with
$\sigma^{1d}_{lim}=400\kms$ halos are acceptable, though a smaller or
larger value of $\sigma^{1d}_{lim}$ may also be tolerated by the
clustering observation alone.

The redshift evolution of the halo-halo correlations for SCDM and LCDM
is plotted in Fig.3b. It is interesting to see that the amplitudes of
the correlation functions do not significantly evolve with redshift,
having only a slight increase with redshift. This is because the
clustering in the mass distribution of dark matter $\xi_m(r,z)$ always
increases with time, or decreases with redshift, but the bias factor
$b(R,z)$ increases with redshift. The two effects seem to be balanced
by each other, giving a very slowly varying of $\xi(r,z)$ with $z$.

 The $z$-evolution of quasar clustering has been studied for more than one 
decade (e.g. Chu \& Fang 1987, Shaver 1988) and the results are quite
scatter.  Some showed a weak decrease of the correlation amplitudes around
redshift 1.5 on scales of $>10\ h^{-1}$Mpc (Mo \& Fang 1993; Komberg,
Kravtsov  \& Lukash 1994; Croom \& Shanks 1996.) Some concluded no
significant $z$-dependence from $z \sim 1.5 - 2.9$ (Zitelli et al. 1992.)
And some even show very weak increase with redshift. Recently,
a weak $z$-increasing correlation from $z < 1.4$ to $>1.4$ is reported 
(Franca, Andreani \& Cristiani, 1997.) 
Obviously, these diverse data cannot provide a concrete test on the 
prediction of Fig.3b.  Although there is quite 
a bit uncertainty in the observed redshift evolution, the current result -- 
no significant evolution of either $z$-increase and $z$-decrease -- is 
consistent with the developed bias models. 

\section{Conclusions}

We showed that velocity-dispersion-selected halos are a possible
mechanism for the bias of quasars. The majority of quasars at redshift
$z\sim 1 - 5$ formed in the environment of new born collapsed halos
with 1-D velocity dispersion $\sigma^{1d}_v \sim 400 \kms$.  Both the
harboring coefficient $f$ per halo and the lifetime of quasars are
$z$-independent. With this bias model, the popular structure formation
models, like SCDM and LCDM, can be fairly well reconciled with data of
the abundance and correlations of quasars at $z \geq 0.5$.

It is interesting to point out that the velocity dispersion identified
halos generally don't have the same mass. Eq.(2) shows that for a
given $\sigma^{1d}_v$, the redshifts the higher, the mass of the halos
the smaller. This result has already been recognized in an earlier study,
which shows that in order to fit with quasar abundance at high
redshift, the mass of the halos has to be smaller than at the lower
redshift (Bi \& Fang 1997).

With this model, one can predict that 1. The environment for quasars
at redshifts from $z \sim 1$ to 5 should be characterized by a
velocity dispersion, $\sigma^{1d}_v \sim 400 \kms$; 2. The amplitudes
of quasar two-point correlation function at high redshifts don't
significantly evolve with redshifts. In the paper, only the models of
the SCDM and LCDM are considered. We can expected that with better
data of quasars becoming available, the bias model of quasars will
play more important role for discriminating among models of structure
formations.

\acknowledgements

We would like to thank an anonymous referee for a detailed report
which improves the presentation of the paper. YPJ gratefully 
acknowledges the receipt of a JSPS postdoctoral fellowship.


\begin{figure}
\epsscale{1.0} \plotone{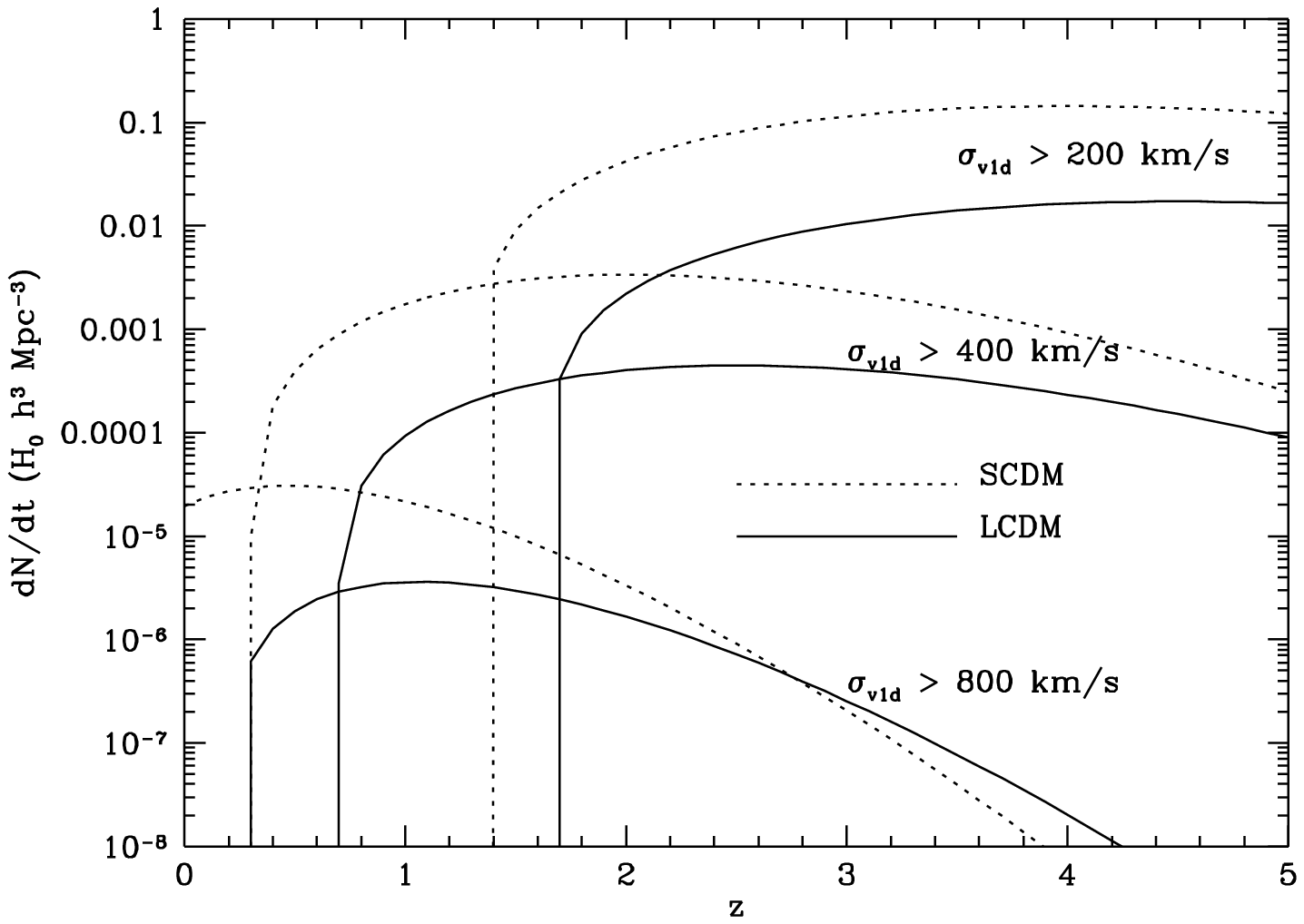}
\figcaption{ The birth rate of halos with the 1-D velocity dispersion
  larger than $\sigma^{1d}_{lim}$. From top to bottom curves,
  $\sigma^{1d}_{lim}$ are $200\kms$, $400\kms$, and $800\kms$
  respectively.  }
\label{fig1}
\end{figure}

\begin{figure}
\epsscale{1.0} \plotone{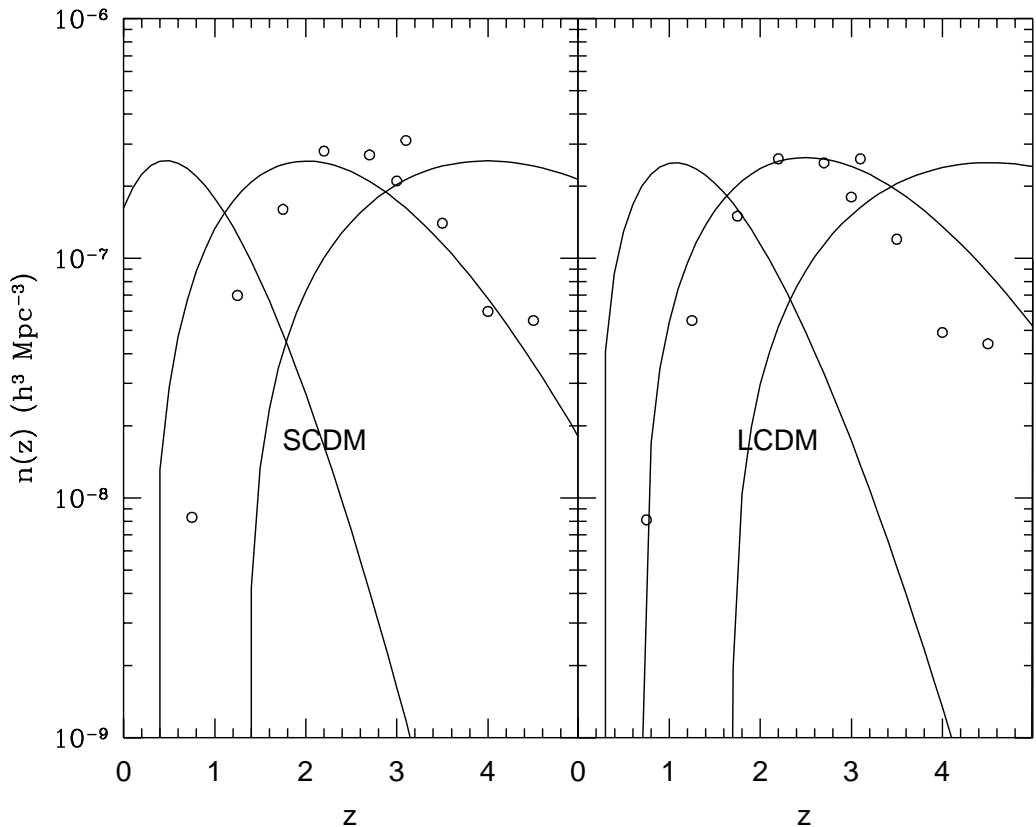}
\figcaption{ The quasar mean density predicted by the model (solid curves)
  versus the observation data (dots; see text). The curves from right
  to left are for $\sigma^{1d}_{lim}= 200\kms$, $400\kms$, and $800\kms$
  respectively.  The mean lifetime $t_{qso}$ for quasar is taken to be
  $10^{7}
  yr$. The harboring rate $f$ is adjusted to match the theoretical
  maximum density to the observed one. In the increasing order of
  $\sigma^{1d}_{lim}$, $f$ is $0.45\times 10^{-3}$, $0.19\times 10^{-1}$ and
  $2.1$ for the SCDM, and $0.84\times 10^{-2}$, $0.33$, $38.6$ for the 
  LCDM.}
\label{fig2}
\end{figure}

\begin{figure}
\epsscale{1.0} \plotone{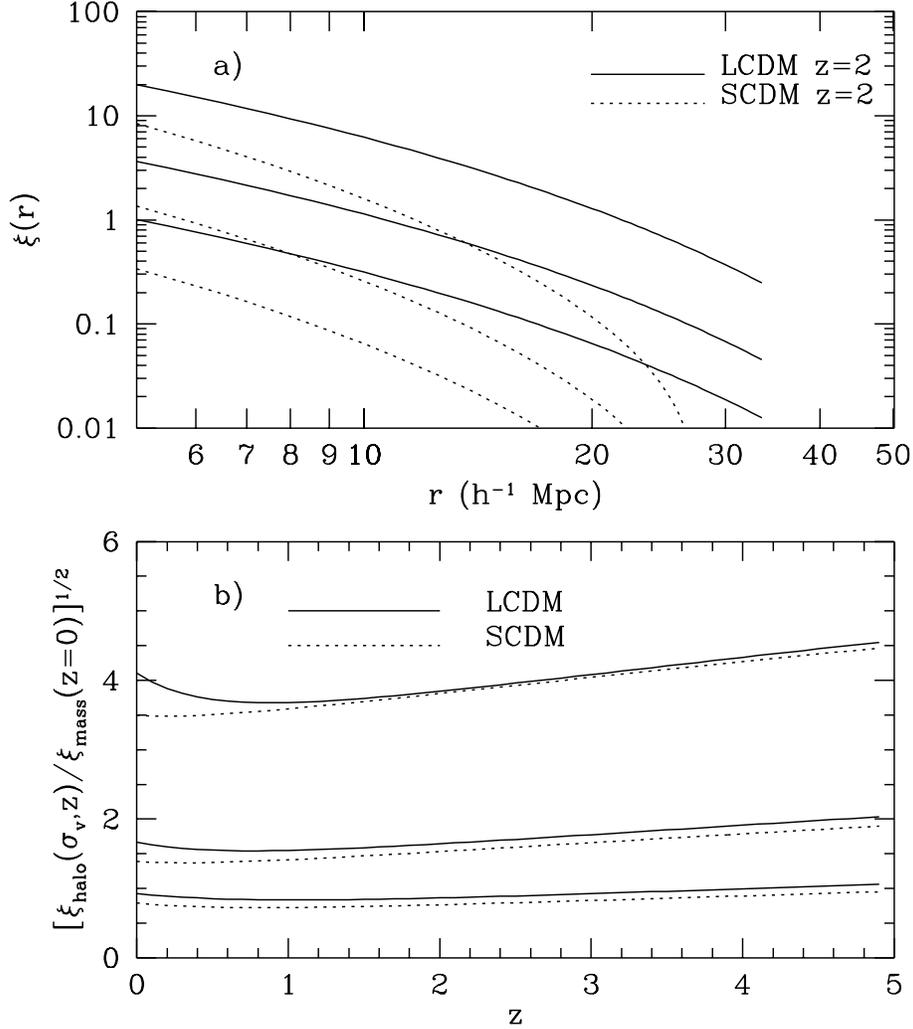}
\figcaption{ a) The correlation function 
  at $z=2$ of the halos with the 1-D velocity dispersion equal to
  $\sigma^{1d}_{lim}$. The curves from bottom to top are for 
  $\sigma^{1d}_{lim}=200\kms$, $400\kms$, and $800\kms$; ---b) 
The evolution of the halo correlation function on the linear 
  scale. The halo correlation function is normalized by the mass
  correlation function at $z=0$. The 1-D velocity dispersion for the halos 
  is taken to be $200\kms$, $400\kms$, and $800\kms$ (curves from
  bottom to top).}
\label{fig3}\end{figure}

\end{document}